\documentclass[preprint,journal]{vgtc}       

\usepackage{mathptmx}
\usepackage{graphicx}
\usepackage{times}
\usepackage[utf8]{inputenc} 
\usepackage[T1]{fontenc}
\usepackage{textcomp}
\usepackage{gensymb}
\hypersetup{
  pdfauthor = {Philipp Geuder, Marie Claire Leidinger, Martin von Lupin, Marian Dörk, Tobias Schröder},
  pdftitle = {Emosaic: Visualizing Affective Content of Text at Varying Granularity},
  pdfkeywords = {information visualization, text and document data, color perception, sentiment analysis, digital humanities},
  colorlinks=true,
  linkcolor= black,
  citecolor= black,
  pageanchor=true,
  urlcolor = black,
  plainpages = false,
  linktocpage
}

\onlineid{0}

\vgtccategory{Research}

\vgtcinsertpkg

\title{Emosaic:~Visualizing~Affective~Content~of~Text~at~Varying~Granularity}

\author{Philipp Geuder, Marie Claire Leidinger, Martin von Lupin, Marian Dörk, and Tobias Schröder}
\authorfooter{
\item
  Philipp Geuder, Marie Claire Leidinger and Martin von Lupin were students of interface design at University of Applied Sciences Potsdam.
  
\item Marian Dörk and Tobias Schröder
  are research professors at the Urban Futures Institute of University of Applied Sciences Potsdam. 
}

\shortauthortitle{Geuder \MakeLowercase{\textit{et al.}}: Emosaic}

\abstract{
This paper presents \emph{Emosaic}, a tool for visualizing the emotional tone of text documents, considering multiple dimensions of emotion and varying levels of semantic granularity. 
Emosaic is grounded in psychological research on the relationship between language, affect, and color perception. 
We capitalize on an established three-dimensional model of human emotion: 
valence (good, nice vs. bad, awful), arousal (calm, passive vs. exciting, active) and dominance (weak, controlled vs. strong, in control).
Previously, multi-dimensional models of emotion have been used rarely in visualizations of textual data, due to the perceptual challenges involved. 
Furthermore, until recently most text visualizations remained at a high level, precluding closer engagement with the deep semantic content of the text.
Informed by empirical studies, we introduce a color mapping that translates any point in three-dimensional affective space into a unique color.
Emosaic uses affective dictionaries of words annotated with the three emotional parameters of the valence-arousal-dominance model to extract emotional meanings from texts and then assigns to them corresponding color parameters of the hue-saturation-brightness color space.
This approach of mapping emotion to color is aimed at helping readers to more easily grasp the emotional tone of the text. Several features of Emosaic allow readers to interactively explore the affective content of the text in more detail; e.g., in aggregated form as histograms, in sequential form following the order of text, and in detail embedded into the text display itself. Interaction techniques have been included to allow for filtering and navigating of text and visualizations. 
}


\teaser{
	\centering
	\includegraphics[width=1\linewidth]{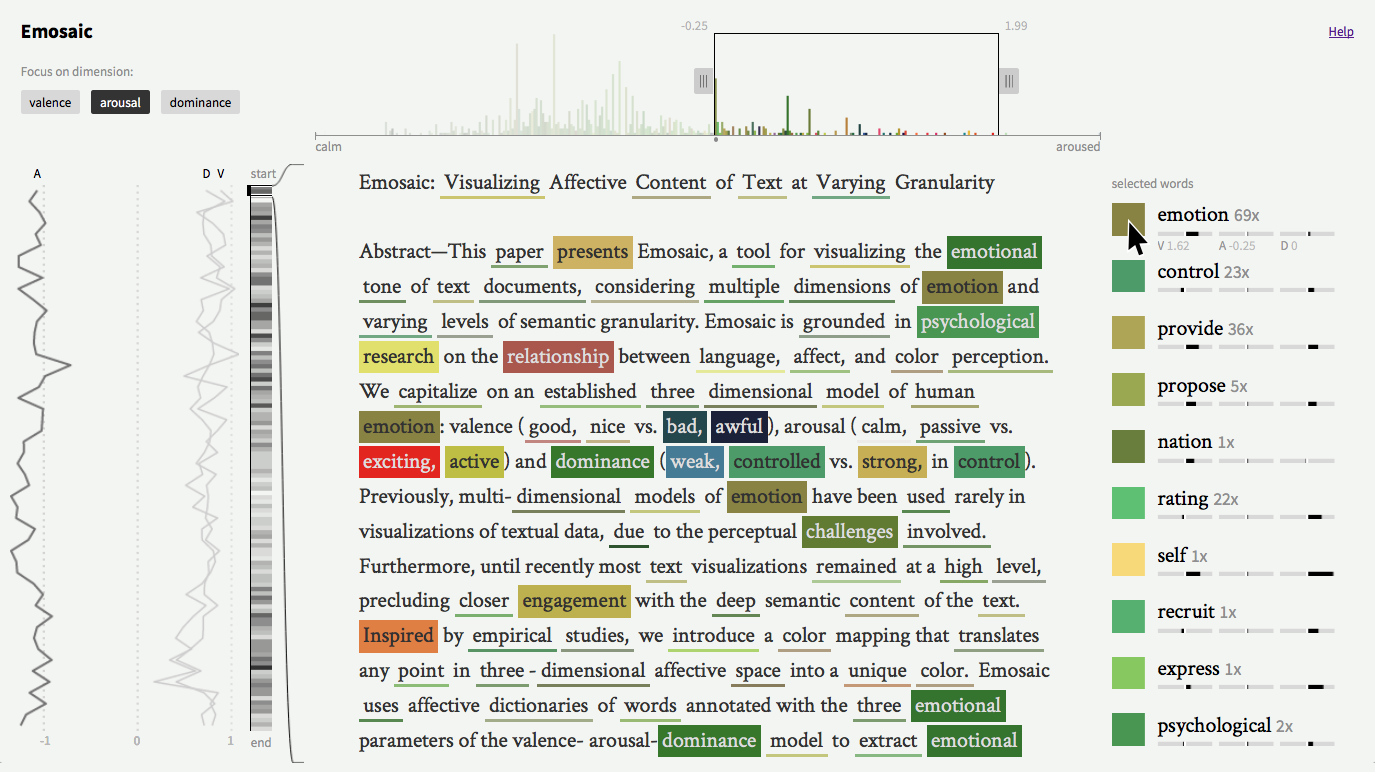}
  \caption{Emotion visualization of this paper with a selected arousal scope of [-0.25, 1.99] while hovering over the word ``emotion''.}
	\label{fig:views}
}

\begin{document}


\firstsection{Introduction}

\maketitle

Human language involves affective connotations.
While certain words are strongly connected with positive or negative emotions, others might trigger more nuanced emotional responses.
With the increasing amount of available text through digitization of printed documents, increasing online communication, and the rise of social media, the automatic analysis of text has become its own research area.
While traditional ways of reading and interpreting texts do not scale to these growing quantities of textual data, quantitative approaches promise new analytical methods for large text corpora.
Pursuing this argument, Moretti~\cite{Moretti2013} coined the term ``distant reading'' to refer to a macroscopic approach to literary analysis.
The thesis is as pragmatic as it is provocative: when considering entire literary eras, it is hardly feasible to study the individual texts in detail, thus we should embrace automatic means to extract high-level semantic and stylistic patterns over thousands of works.
With the present research, we seek to contest such a binary contrast between distant and close analyses of text, in particular with regard to its emotional tone.

Several automatic approaches have been proposed to extract the emotional content of text~\cite{Liu2012,Pang2008}, commonly referred to as sentiment analysis or affect detection.
These methods focus on opinions conveyed in text such as reviews~\cite{Turney2001}, on general affective tonality, or on identifying expressions of specific emotions such as happiness, sadness, or anger.
The majority of such studies have relied on rather simplistic models, either representing human emotions with a restricted set of labels that are easy to understand, but lack emotional variety, or reducing emotion to a one-dimensional spectrum ranging from good to bad (or happy to sad).
In contrast, more sophisticated models of emotion capture a variety of emotional concepts that can be qualified along multiple dimensions such as valence, arousal, and dominance~\cite{Bradley1999,Osgood1975,Schmidtke2014}.
Even so, visualizations of sentiments in text have focussed on one-dimensional models of emotion and mostly remain in a distanced perspective on the textual documents~\cite{Zhao2014}.
More generally, visual analyses of text documents in support of distant and close reading do not need to pose contradictions~\cite{Koch2014}.
In sum, recent research underlines the opportunity to visualize textual data along multiple dimensions of emotion and at varying levels of detail.

The aim of our research is to develop an approach to text visualization that reveals nuanced differences in emotion at multiple levels of granularity. The main challenge is to integrate psychological models of emotion with interactive text visualization techniques supporting both distant and close readings. Based on empirical associations between language and emotion, we propose a word-to-color mapping that can be used to visualize and read text documents along valence, arousal, and dominance (VAD). We map these emotion dimensions to three color parameters, resulting in a mapping with a unique color for each emotional nuance. Thus we try to combine the visual simplicity that specific emotion labels provide with the emotional complexity that is latent in our language. The three independent color parameters of the HSV (hue, saturation, and value) color space provide a basis for the differentiation of the three emotion parameters. When the color mapping is integrated into the text display, we aspire to make visible how emotions and their components develop through a textual narrative and how words are emotionally connected. The emotional patterning of the text becomes an additional structure that can be interpreted. We thus explore how the emotional vicinity of words can be used to ‘read’ the text in a non-linear fashion. To test these ideas and aspirations, we introduce \textit{Emosaic}\footnote{\url{https://emosaic.de}}, a web-based visualization tool that implements the word-to-color mapping and supports multiple levels of emotion analysis of text documents. 
\section{Related Work}

Our research on visualizing emotion in text relates predominantly to
the analysis of large document collections at different scales, emotion research in general, 
the relationship between color and emotion, and the combination of color and text to visualize emotion.

\subsection{Visualizing data at different scales}

There is a considerable body of research on visualization techniques for analyzing large amounts of data at varying levels of granularity and detail.

Interface schemes that allow users to move between more detailed views of a dataset to an overview and steps between have been reviewed by Cockburn et al.~\cite{Cockburn2008}.
They particularly summarized and categorized interaction techniques that are relevant for browsing text documents,
suggesting that people need a way of changing what is being displayed in order to make sense of the data.
Visual abstractions of text documents that help the user explore and analyze a large amount of text have already been presented by Koch et al.~\cite{Koch2014} with their ``Varifocal Reader'',
which provides a text visualization at multiple layers displaying different levels of scale and abstraction.
Similarly, it has been shown how such multi-level text visualizations can support the analysis of tagged documents~\cite{correll2011exploring} and examining the distribution of topics across a text corpus~\cite{alexander2014serendip}.
All views are linked to each other, so the user can navigate through the visualization while tracing the current position across all aggregation levels.
Koch et al. divided their views in chapter, subchapter and lines, thus different detail levels of a book can be examined.
We also aim to provide an interface that lets readers compare findings or test hypotheses along different abstraction levels.
However, we focus on the affective quality of a text and therefore on providing different basis for abstractions and explorations.
We hope to contribute to a deeper understanding of emotionality in texts by visualizing connections that would not be visible without the color mapping we developed.
Dörk et al.~\cite{Doerk2012} proposed a tool for exploring comprehensive information spaces
containing multiple facets and relations.
In using a visualization with semantic and structural relations to connect them, the visualization creates context and reveals hidden aspects that would not be detectable when looked at them individually,
which resonates with our aim to reveal relationships of words or aspects of the texts that would not be discovered by simply reading the text. 
These ideas were further explored in a text visualization
that did not rely on any faceted structures,
but on proximal occurrences of words in unstructured texts~\cite{Doerk2015}.
With aim continue this line of research,
by enabling the analysis of emotional relationships among words in a document,
while still encouraging the close engagement with the actual text and its words.

\subsection{Words and colors of emotion}

There have been decades of research on affective meanings of linguistic content, which our present work builds upon. There is not one, but two significantly different models to represent emotions: the categorical model and the dimensional model.
The former is represented by Ekman~\cite{Ekman1992} who introduced a concept of six basic emotions: anger, disgust, fear, joy, sadness and surprise. 

The latter goes back to much earlier studies by Osgood et al.~\cite{Osgood1957},
during which people were asked to rate words on a broad variety of bipolar scales. 
They discovered that three main factors accounted for most of the variation in the data: the evaluative factor (also known as valence or pleasure), the activity factor (also known as arousal) and the potency factor (also known as dominance).
These dimensions are known today as basic constituents of human emotion, 
regardless of whether measured with linguistic methods or nonverbal signs and physiological symptoms of emotion~\cite{Fontaine2007}. 
The fact that these three facets of emotions can be detected from linguistic material in virtually all human languages~\cite{Osgood1975} reflects the fundamental role of affect in the regulation of human communication, from basic nonverbal dynamics to complex cultural construction of meaning systems~\cite{Rogers2014,Thagard2014}. 
Consequently, studies have found substantial agreement among members of one culture about affective meanings of concepts~\cite{Ambrasat2014,Heise2010}, providing justification to our approach, 
which treats emotions as properties of text rather than idiosyncratic reactions of readers.  
These studies provide one important empirical foundation for our research
on visualizing emotion in text.

Color is an important visual attribute for text visualizations and is often used to display different textual features.
E.g., color has been used to highlight classification of words or phrases~\cite{Koch2014} or to mark words in parallel texts~\cite{Jong2009}, and to represent the phonetic and prosodic qualities of the text~\cite{clement2013sounding}.
However, the color mappings in these kinds of visualization tend to be relatively arbitrary.
In contrast, Setlur et al.~\cite{Setlur2016} propose the usage of meaningful color associations in visualizations in order to ease the associations to a given color and what they stand for.
They developed a system that determines if a term has a strong association to color, then assigns a semantic color to it using the basic colors associated with the term along with further linguistic analysis like semantic context.
We conjecture that empirically meaningful color assignment increases the readability of a data visualization by making the encoding of color easier to discover and to remember.
Even though Setlur et al. provide such a method in principle, their approach is only suitable for visualizations that depict a limited number of categories.
Our aim is to develop a system that depicts nuances of an emotional spectrum through color, thus a representation of tendencies, slants, and vicinity in a dense multidimensional space.
We do not map discrete colors to specific words,
but we rather map their underlying affective connotations to color parameters.
Thus, we associate the affective meaning of text with the color spectrum itself. 

Most of the earlier work dealing with the emotionality of color fails to control other color parameters than hue. 
Valdez and Mehrabian~\cite{Valdez1994} studied emotional reactions to color hue, saturation, and brightness using the pleasure-arousal-dominance model, which largely corresponds to the VAD model.
They found out that color preferences or color emotion association regarding hue tends to be weak. 
However, brightness and saturation receive a stronger emotional associations (more saturated colors elicit greater levels of arousal, brighter colors appear more pleasant).
We base our color mapping on this research.

\subsection{Visualizing emotion through color}

There already have been several approaches to visualize
the emotional content of text through color,
interestingly most of them were done in the context of social media.
``We Feel Fine'' by Kamvar et al.~\cite{Kamvar2011} performs sentiment analysis on the content from web blogs and online communities. 
In this experimental and artistic visualization,
emotional analysis of blog postings is combined
with additional information such as demographic and weather data. 
While each circle represents a specific emotion word,
there are only a few colors representing emotional categories and do not incorporate the more nuanced notion of emotion of the dimensional model. Furthermore, the visualization is not geared for the interpretative analysis of larger documents. 

Happy emotions are represented by a bright yellow, anger is represented by red. 
While this color coding helps the user to differentiate between different types of emotions, it reduces the spectrum of colors to very few spots in the overall emotional space.

A similar web tool is ``We feel'' by Milne et al~\cite{Milne2015}.
It analyzes English tweets and maps them to a wheel of emotions, a streamgraph, and a globe.
The goal behind this project is to record the dominant collective sentiment of a country at a certain time.
Similarly, the color coding is based on the six primary emotions of Parrott~\cite{Parrott2001}: love, joy, surprise, anger, sadness and fear. The words of the tweets matching these categories are assigned a color.
The tool also visualizes secondary emotions derived from the six basic emotions. 

PEARL, a visual analysis environment for examining tweets, by Zhao et al~\cite{Zhao2014}
is very related to our work here.
It's a multi-view interface, but it is primarily 
based around a timeline visualizing the emotional development in a tweet corpus.
While the tool does expose display VAD values in scatterplots and statistical tooltips,
the colors used in the visualization correspond to emotional categories.

The main limitations of existing visualizations of emotion in text are twofold:
first, the three recognized dimension of emotions have not been utilized for
finding a color mapping that can be used to visualize and explore text documents;
second, the visualizations are often disconnected from the underlying text,
making the shift between broad analysis and detailed examinations of a text difficult.
Our intention is to develop a more nuanced word-to-color mapping
and find ways that better integrate different levels of analysis.
\section{What is the Tone of this Text?}

Emotions are complex and subjective, and have no clear semantic visual analogy, yet they play a major part in language and writing. While categorical labels for a few basic emotions are not suitable for representing the full spectrum of emotions,  multidimensional models of emotion are a promising way forward in the sentiment-aware analysis of text. However, so far no suitable visualization of multidimensional emotion models have been proposed. A visualization of texts in a three-dimensional space would decontextualize the text’s structure, which is essential for the close engagement with the text. Furthermore, three-dimensional visualizations are known to cause perceptual problems due to occlusion and perspective distortions. Yet, the three-dimensional VAD model provides an empirical source for ascribing nuanced emotional information to words, which could help the analysis of this latent layer of emotion in human communication, in particular in text.
Generally, we are interested in finding a way to represent the three-dimensional model of emotion interwoven into the text in a way that encourages its reading along multiple levels of interpretation. 
More specifically we set ourselves the following three goals:

\begin{enumerate}
\item \textit{Represent emotion as color.} We aim to devise a mapping of emotional content to color parameters for every numeric value of the three emotional dimensions of valence, arousal, and dominance.
\item \textit{Support analysis at multiple levels.} We wish to create visualizations that enable the open-ended exploration and interpretative analysis of emotional content in text at different levels of abstraction.
\item \textit{Support new readings along sentiment.} We hope to open up new opportunities for reading and interpreting large documents with particular attention to emotions, opinions, and sentiments.
\end{enumerate}

\section{Mapping Emotion to Color} 

Nearly all research on emotion refers to the VAD emotion model that divides emotions into three components: valence (the pleasantness of a stimulus), arousal (the intensity of emotion provoked by a stimulus), and dominance (the degree of control exerted by a stimulus).
The ANEW norms by Bradley and Lang~\cite{Bradley1999} are a set of normative emotional ratings based on the VAD model.
They collected ratings for 1,034 English words using the Self-Assessment-Manikin (SAM), an affective rating system, originally devised by Lang~\cite{Lang1980} that uses pictograms to record the dimensions of affective reactions. 
They carried out three kinds of ratings, largely within the scope of Osgood, Suci and Tannenbaum's theory of emotions (for critique and comparison, see~\cite{Schmidtke2014}).
The first type of rating concerns the valence (or pleasantness), the second one the arousal and the third one the dominance of the emotion.
However, such a small corpus of words is too small to analyze comprehensive text documents.

Warriner et al~\cite{Warriner2013} extended Bradley's database and collected numerical ratings for 13,915 English lemmas that include stimuli from nearly all categorical norms. 
Although they did not incorporate SAM, but used numerical ratings and let users rate words on one single dimension (in opposite to the ANEW database where participants rated words on all three dimensions). 
Warriner et al posit that the methods are roughly equivalent, because their ratings correlated highly with the ANEW ratings.
The participants who were recruited via Amazon Mechanical Turk had to rate on a scale from 1 to 9 how they felt while reading the words.
The scale ranged from 1 (happy or excited or controlled) to 9 (unhappy or calm or in control), with 5 being the numerical value for a neutral feeling.
The valence and arousal ratings were retroactively reversed, because higher numbers usually go with positive anchors (according to Rammstedt \& Krebs~\cite{Rammstedt2007}) leading to a scale of valence / arousal / dominance from the minimum 1 (unhappy / calm / controlled) to the maximum 9 (happy / excited / in control).
We use Bradley's ratings in our color mapping to utilize the largest database currently available, to our knowledge. 

\subsection{HSV color space}

Representing emotions with color requires mapping the dimensions of the VAD model to an appropriate model of the color space. As we aim to help users make sense of the emotional composition of texts, the color mapping should be comprehensible and decodable. After short iterations with different three-parametrical color systems, we opted for the HSV color space as the most appropriate one for mapping emotion to colors. HSV has highly independent color parameters and is particularly similar to our human understanding of color composition. The characterization of a color position in the HSV color space is defined by the following parameters:

\begin{itemize}
	\item Hue \emph{H}: dominant wavelength in the spectrum, as a color angle on a color circle [0, 360], e.g., 0° for Red, 120° for Green, 240° for blue.
	\item Saturation \emph{S}: ``colorfulness'' or ``pureness'' in percentage [0, 100].
	\item Value \emph{V}: ``brightness'' or ``lightness'' in percentage [0, 100].
\end{itemize}

We assigned valence to hue, dominance to saturation, and arousal to brightness. This conforms to the hypothesis by Valdez and Mehrabian~\cite{Valdez1994} who studied the emotional reactions to color regarding, hue, saturation and brightness. 
They found out that more saturated color elicited a greater level of arousal and brighter colors a greater level of pleasantness. 
There were weaker results relating hue to emotional reactions.
However, there has been the most evidence of a dependence of valence to hue compared to the effect of arousal or dominance to hue.
Even though the results in this research of relating emotional reactions to colors have been rather weak, we choose to assign the parameters of the VAD-model to the color parameters of HSV accordingly.
Valence is assigned to hue, arousal is assigned to saturation and dominance is assigned to brightness.

We base this decision on the fact that the classification of VAD to HSV based on Valdez is only semantically relevant for us, in terms of facilitating to reading of the encodings.
However, we acknowledge that such a mapping does not do full justice to the nuance and richness in the emotional space. Instead the mapping serves as a broad heuristic for the analysis of text. The assumption is that the close reading of text likely compensates for inaccuracies. Furthermore, we do not aim to create a semantic translation of emotion to color (like e.g. Lindner et al)~\cite{Lindner2012} or depict the associations between color samples and words (see Wexner, 1954~\cite{Wexner1954}.
Instead, we establish a general mapping between colors and words according to their position in the VAD space.

\subsection{Emotion to color translation}
\begin{figure}[htbp]
    \includegraphics[width=\columnwidth]{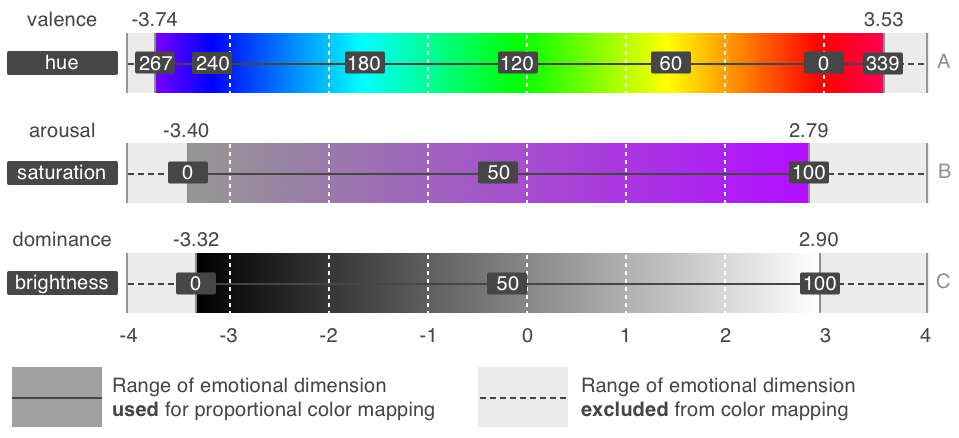}
    \caption{Specific color mapping for the emotional dimensions (A) valence, (B) arousal and (C) dominance. The ranges that have been used and excluded for the proportional mapping of the respective color parameter are emphasized.}
    \label{colorcircle}
\end{figure}

To emphasize the value of emotion ratings, we used a scale that supports the quick understanding of low, neutral and high numbers, while keeping the size of the range comparable. Consequently, we mapped the numbers from a range of 1 to 9 to a range of -4 to 4. All following charts and numbers are converted to this scale, including the results from Warriner et al.~\cite{Warriner2013}. While the possible range of the rating system is from -4 to 4, the actual minima and maxima of the ratings, however, lie in a smaller range:

\begin{itemize}  
\item -3.74 (``pedophile'') to 3.53 (``vacation'') for valence 
\item -3.40 (``grain'') to 2.79 (``insanity'') for arousal
\item -3.32 (``dementia'') to 2.9 (``paradise'') for dominance 
\end{itemize}
We used these minima and maxima of the database for our mapping to ensure the applicable extent of the color spectrum (see Figure~\ref{colorcircle}).
While saturation and brightness have clear minima and maxima and can therefore easily be mapped, hue is defined by a continuous color gradient. 
To avoid approximation of the color hues at the marginals of the color translation scales, we consciously left out a part of the color gradient (about a fifth of the circle, namely 72°) resulting in a clear definition of minima and maxima.

Purple equates to the minimum, red to the maximum and green to the middle of the valence scale.
We base this on the word ``love'' which has a valence-rating of 3. 
The color gradient has been modified in a way that love definitely equates to red, because of a strong correlation of the word and the color in Western culture. 
We are aware that there are more possibilities for an assignment (which we will discuss later in this paper), nevertheless it has been proven as easily decodable and therefore a suitable anchor in the color mapping.

This is the color mapping, we propose (see Fig.~\ref{colorcircle}):

\begin{itemize}
\item valence to hue : [-3.74, 3.00] to [267, 0]\\ and [3.00, 3.53] to [360, 339]

\item arousal to saturation: [-3.4, 2.79] to [0, 100]

\item dominance to brightness [-3.32, 2.9]  to [0,100]
\end{itemize}

All emotional parameters are linearly mapped to the color parameters. 
There have been attempts to adjust the color space according to the research results of Valdez~\cite{Valdez1994}.
As mentioned before she found out that brightness and saturation have a bigger effect on emotion:
Brighter and more saturated colors are more pleasant,
less bright and more saturated colors are more arousing,
and less bright and more saturated colors induce greater feelings of dominance.
 In combination with the findings of one of her studies, the resulting relationship between the emotion and color dimensions could be calculated in the following ways:\\

Pleasure = .69 Brightness + .22 Saturation\\ \vspace{-0.075in}

Arousal = -.31 Brightness + .60 Saturation\\ \vspace{-0.075in}

Dominance = -.76 Brightness + .32 Saturation\\

\noindent Solving these equations for Brightness and Valence, we get:\\

B = 1.26v - 0.46a\\ \vspace{-0.075in}

S = 0.64v + 1.43a\\

However, these adaptations to our color mapping result in colors, where the single color parameters are harder to decode. Since our goal is to create a color mapping that allows for a simple reading of the color into its parameters, we maintain the linear mapping, solely because it offered better results concerning that matter. 
As the database by Warriner et al.~\cite{Warriner2013} is limited to 13,915 English lemmas, the mapping mau not apply to every word of a user-generated text. We will refer to words with affective ratings as “emotionally relevant”.

\section{Design}

The proposed mapping has been implemented in the context of a web-based reading tool \textit{Emosaic} , which allows the exploration and inspection of user-defined textual content along the VAD emotion dimensions. The interface can be regarded as a first prototype to evaluate the introduced emotion-to-color mapping for interpretative analysis and exploration of affective text content. Besides the integration of color, other levels of abstraction have been applied to facilitate multi-level readings of the emotional tone of a text and also to support the comprehension of the underlying cognitive methodology. The tool allows browsing through a text using semantic cues, while color indicates affective content. Words can be highlighted based on emotional dimensions, to inspect desired passages regarding specific emotional tones. Inside the user interface, varying granularities are assigned to several views in order to support both close and distant reading.

\begin{figure}[htbp]
    \includegraphics[width=\columnwidth]{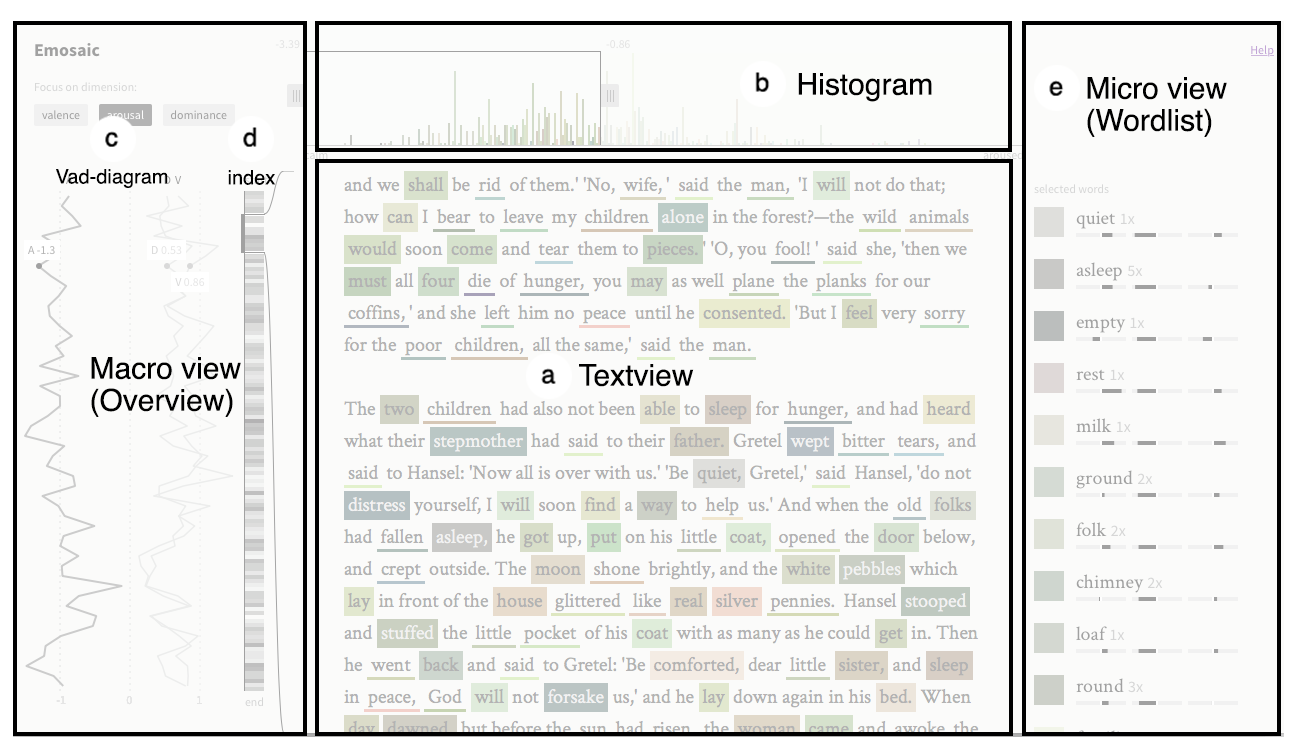}
    \caption{The different parts of the Emosaic interface}
    \label{interface}
\end{figure}

Within the \textit{text view} the text is presented inside a scrollable window, allowing for a familiar reading experience, while emotionally relevant words are highlighted with the respective color according to their position in the VAD space (see Fig. \ref{interface}a). 
A \textit{histogram} presents the distribution of each emotional dimension (see Fig. \ref{interface}b).
Furthermore it controls the selection and highlighting of those relevant words.
An overview of the emotional tone along the text is provided as well as an indication of word occurrences, depending on user selections (individual words or ranges in the histogram). The left side of the interface provides a \textit{macro view} on the text (see Fig. \ref{interface}c,d). 
Inside the interface, either single words or words sharing emotional similarities can be selected. 
This selection is the basis for a detailed view, where the composition of emotional dimensions can be inspected. 
The right side of the interface offers a \textit{micro view}, where words that are near the current selection with thin the emotion space are displayed and linked to other text passages (see Fig. \ref{interface}e). 
The macro and micro view can be used for quick navigation to desired text passages. 
To accomplish fluid interactivity all the views are tightly coupled.

\subsection{Use of semantic color coding}

The ability of the human perception system to distinguish colors and detect similarities not only in hue, but also in saturation and brightness is a fundamental basis of our approach and thus heavily used inside the user interface. 
Every word that is emotionally relevant has an affective characterization (valence, arousal, dominance) and thus a respective color (hue, saturation, brightness) attached. 
Selecting words also results in the selection of emotions and vice versa.
The color that is attached to a word is always apparent either as underlining or background. 
Words that share similar values along all three emotional dimensions have a related color attached. 
Differences in dimensions result in changes of the corresponding color parameter. 
The permanent visibility of the semantic colors make steady comparisons possible.
To avoid confusion, throughout the interface color is only used to indicate affective content that contains the three emotional dimensions. 
Other elements like buttons or labels have neutral tones such as gray or black.

\subsection{Visualizations of varying granularity}

The Emosaic interface features four distinct visualizations that all comprise visual representation and interaction capabilities in support of the interpretative analysis of text at different levels of abstraction.

\subsubsection{Text view}
The textual basis of the tool is the user input of plain text. 
To enable the experience of close reading the readability of the inserted text remains, including the original punctuation as well as line breaks of headlines or paragraphs.
The text is made available within a continuous scrolling window, allowing the reader to peruse even large documents from the beginning to the end without any more interaction than scrolling.
Emotionally relevant words are indicated by color-coded underlining or background, depending on the state of selection. 
There are three states: 

\begin{enumerate}  
\item a single word is selected leading also to the selection of a single emotion, 
\item multiple words are selected, based on shared one-dimensional parameters of emotion, or 
\item no word is selected. 
\end{enumerate}

Selection is indicated by a color-coded background that functions as highlight, while unselected words are less salient being only underlined with their respective color. Since the readability remains, clusters of similar emotions inside a passage can be spotted easily. The visual complexity could become quite high when too many words are inside the visible area, leaving the user overwhelmed with color representations. This is counteracted by a relatively large font size of 20 pixels. This leads to fewer words within the visible area and thus decreases the visual complexity and density. The aim is also to encourage close reading and to make reading long passages more pleasant. 

The inserted text is preprocessed using lemmatization in order to find corresponding entries inside the used database of lemmas and their emotional composition. 
Good results regarding speed and accuracy have been achieved using the natural language processing library ``spaCy'' ~\cite{spacy}.
The overrepresentation of certain common words in texts results in a distinct peak in the histogram for most texts. Due to the fixed height of the histogram, the rest of the graph flattens and loses detail. To resolve this issue we decided to remove stop words from the list of emotionally relevant words (e.g., ‘be’, ‘do’, ‘have’).
Conceptually, the length of an inserted text is unlimited. To avoid performance issues of browsers, however, there is a limitation of processing currently set to 20,000 instances including punctuation. This is subject to change with computational performance improvements.

\subsubsection{Histogram}

The histogram view displays the distribution of emotionally relevant words along the currently selected emotional dimension.
A button for each emotional dimension is available to switch the displayed parameter as desired.
As a key feature it is positioned in the top center of the interface. In its static form it serves as an overview of the emotional distribution, while also functioning as a selection tool for one-dimensional ranges.

As common for the design of a histogram, the case units (bins) are placed along the x-axis. 
The intention of this horizontal alignment is to build a visual contrast to the vertical aligned diagrams of the macro view that correspond to the text flow.
The histogram adapts to the possible range that results from the database. 
Consequently, every one of the three dimensions has a numerically different minimum and maximum.
This concept was applied to let the reader focus on the relevant range as well as making efficient use of available screen space.
This adaption, however, is not kept from the user, as there is an interactive label containing the current value when hovering over the position of interest.

The histogram has a fixed height. 
The original height of the bin containing most entries is mapped to the defined height, the remaining rectangles are scaled proportionally. 
The color of each bin results from calculating the average of the colors of words assigned to it.
The resulting display is meant to convey the methodology behind the color mapping: the valence histogram depicts a continuous spread of the hue from blue to red, the arousal view shows a spectrum of color from low to high saturation, and the dominance histogram features a color spectrum from dark to bright. These color ranges then also serve as a kind of legend for the color encoding.

Common mouse interactions are applied to select ranges in a quick and reversible manner. When placing the pointer over the histogram, a selection can be made using click and drag interaction. Changing opacity of the unselected area as well as a thin frame surrounding the selected area indicate that a selection has been made. Labels on the edges of the frame communicate the exact range that has been picked. Movable controls are attached to adjust the ends as desired and thereby change the range. Panning interaction can be used to shift the selection frame on the horizontal axis, while the size of the range remains the same. The selection inside the histogram is coupled to the highlighting of words inside the text view. Words inside the selected range become more prominent through background filling. The link between the views also works vice versa, as the position of a word in the histogram is indicated by a circle marker when it is being hovered over in the text view. The selection of a range can be removed either by clicking on the unselected area of the histogram or by selecting a single word in the text or micro view.

\subsubsection{Macro view (text overview) }

Since there is only a part of the text visible in the text view, two juxtaposed visualizations in the macro view support overview and navigation: \textit{vad-diagram} and \textit{index}. The vad-diagram provides visual cues about the trend of each of the three emotional dimensions. The index shows word occurrences of the currently selected emotional scope and the position of the visible passage with reference to the whole text. Clicking on a position of interest leads to a representation of that text passage within the text view.

The \textbf{vad-diagram} consists of vertical line graphs showing the aggregated trend of each emotional dimension along the text individually. This allows the viewer to quickly grasp the emotional slant for different parts of the text and help discover noticeable emotional fluctuations in the language. Since this visualization represents the affective dimensions independently, the line graphs do use the semantic color encoding.
The x-position of a graph’s data point, the one-dimensional value, is based on the average value of a specific part of the text. Therefore, the text is divided into predefined parts depending on the window height. Thus, the y-position of a word in the text is crucial for the assignment of a part.
The width for this visualization is fixed. The edges are defined by the minimum and maximum values among all dimensions inside the present text to make proper use of available screen space. As this results in individual shifts and different scales for each inserted text, orientation guides are provided in form of dotted vertical lines for each full unit. In addition, when hovering over a position of interest, labels are shown providing distinct numerical values for that passage.

The \textbf{index} indicates the occurrences of the selected words across the entire text in a narrow and minimal visualization akin to a one-dimensional heatmap. It is only present while a selection is active. This view is tightly coupled with the selective feature of the histogram as well as the single word selection of the text and micro view. In combination with the word selections, the index represents the distributions of specific emotionally relevant words and particular values or ranges of emotional dimensions along the text.
Flat rectangles that are superimposed upon each other relate to a specific part of the text. A gray-scale filling indicates the amount of currently selected words in that part of the text. While white refers to no appearance, a 80\% black indicates the maximum. 

\subsubsection{Micro view (words lists) }

The micro view contains information about the emotional composition of words of the current selection. A list entry always consists of the word, its count of occurrences in the text, and its assigned color represented by a colored square. Every entry can always be selected to support exploration.
The micro view is based on shared parameters and shows vicinity among emotions in two different manners, depending on the state of selection: selected words and emotionally similar words. In both views, the colors serve as a quick indication to evaluate the degree of vicinity. When a word in the list is selected, the first instance will be shown in the text view. Consequently, the entries can be used to navigate through the text in a nonlinear manner. In contrast to the text view, where adjacent words and colors can be seen in context, the word lists offer a selective view onto affective content and navigational links in a serendipitous, yet comprehensible manner. 

\textbf{Selected words} are listed when a selection has been made along one affective dimension in the histogram. Words are then sorted by the dependent value of one of the three dimensions in ascending order. Since this list type is based on the selection of a one-dimensional range, differences in the other two dimensions are most likely to occur. Those differences can be perceived by their colors and inspected in detail by reading the labels and charts beneath each entry. Each word is accompanied by three juxtaposed horizontal butterfly bar charts showing the composition of the dependent values valence, arousal and dominance. To convey a word’s prevalence in the text, a number showing its count is given as well.

Emotionally \textbf{similar words} that appear in the text, on the other hand, are listed when a single word is selected. Similarity is calculated over all three emotion dimensions. Words that differ by no more than a value of 0.5 in the three-dimensional space are considered similar. This threshold was iteratively tested to be suitable. The word that has initially been selected is presented on top of the list in the same way as the words in the selected words view. Similar words, however, are presented without the charts, as they share nearly the same values over all three dimensions anyway. Almost-equal colors are evidence for the emotional similarity among the words, but fine nuances can be recognized based on tiny variations in their respective color. In contrast to the other list type, the size of this list can not be limited. Some words, especially those with neutral valence, may have long lists of related words. Others, such as those on the edges of the spectrum, are most likely to have less or even no emotional correspondence. Both can serve as meaningful cues about the affective position of a word in the overall text.

\begin{figure}[tbp]
    \includegraphics[width=\columnwidth]{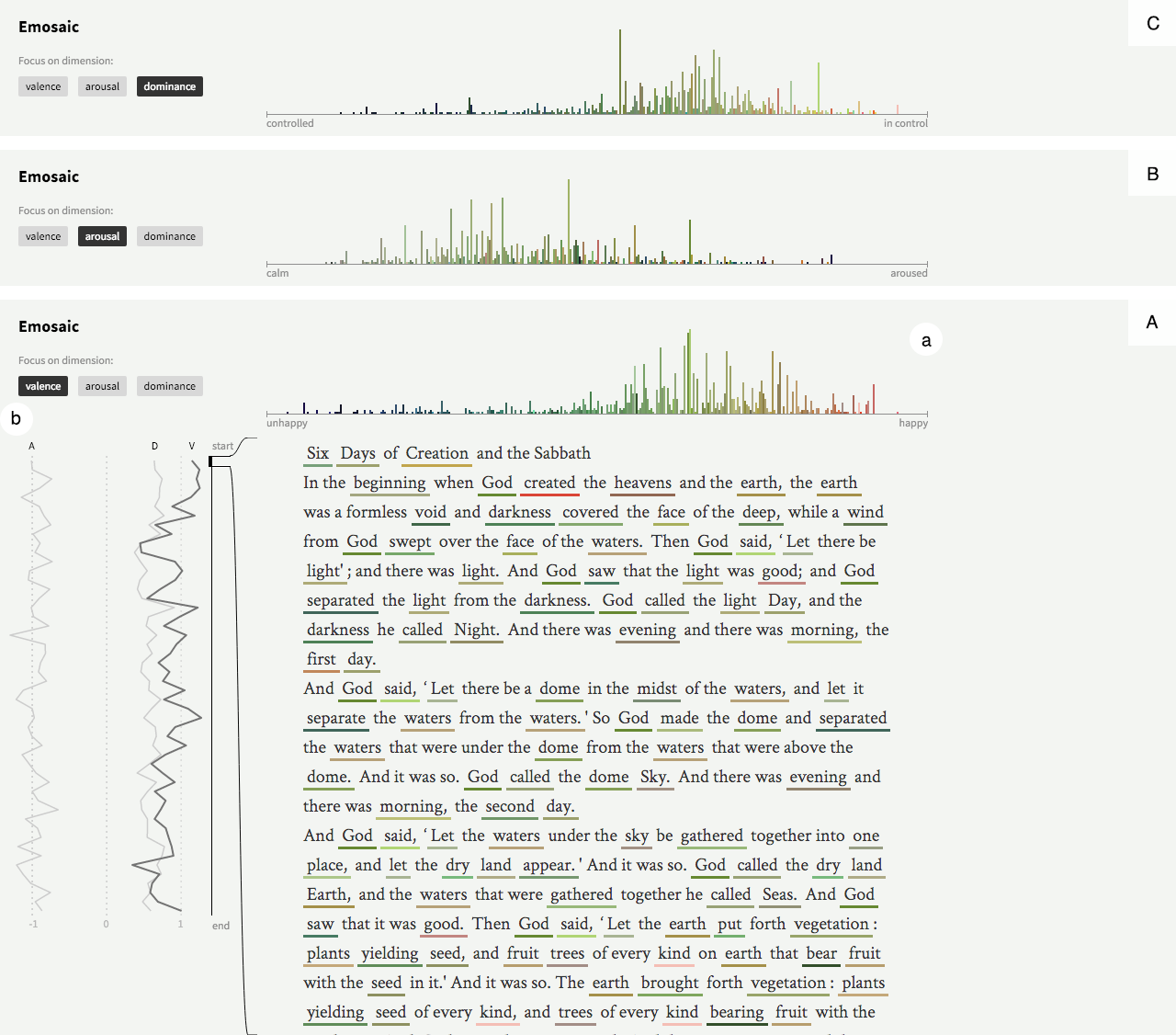}
    \caption{User Interface of \emph{Emosaic} and different views of the Histogram.}
    \label{walkthrough1}
\end{figure}

\section{Walkthrough}

To illustrate the functionality of the color mapping, we present ways of reading the Book of Genesis based on its affective content. The following usage scenarios demonstrate how \textit{Emosaic} invites readers to detect peculiarities in the text through its emotion visualization, which can then be explored further  through studying the text in a combination of close and distant reading. The interface is meant to support different analysis types, each allowing the scholar to compare findings or test hypotheses along different aggregation levels, leading to a deeper understanding of emotionality in texts by visualizing connections that would not be detectable without the color translation and the respective visualization of the text. To gain insights about the emotional content of this text, the viewer inserts it and clicks ``submit'' on the start screen leading to the actual data visualizations.

As already mentioned, the interface is divided into \emph{Text View}, \emph{histogram}, \emph{Macroview} and \emph{Microview}.
Every view provides different information about the emotional content and thus different possiblities for exploring and analysing the text.

\subsection{Examining the general emotional slant of the text}

Valence is selected by default, thus the histogram is arranged along the valence dimension. (see Fig. \ref{walkthrough1}A)
To get a first impression of the text's general affective state, we can take a look at the histogram. (Fig. \ref{walkthrough1}Aa)
The user can see that the general slant of the text should be rather happy than unhappy because most words are assigned to the positive end of the valence scale.
By clicking on dominance and arousal, the histogram changes its parameters and arrangement and the viewer can examine the general slant along those dimensions. (Fig. \ref{walkthrough1}A,B)
According to the dominance histogram, the text seems to contain more words that express being rather in control than controlled, however the biggest peak is at 0, thus neutral.
The arousal histogram shows peaks at the negative end of the scale, thus words that were rated as more calm, than aroused.

\begin{figure}[th]
    \includegraphics[width=\columnwidth]{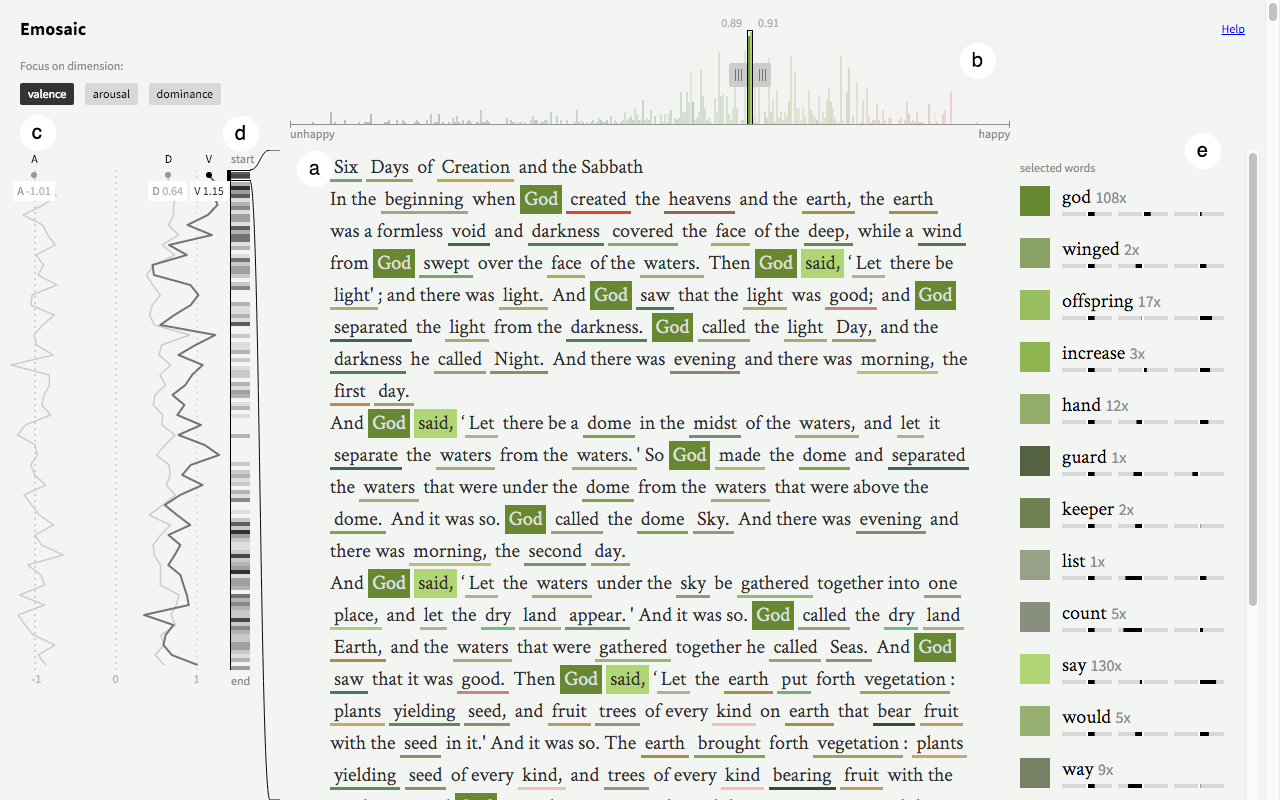}
    \caption{a) Textview, words that are selected are highlighted through a background color, b)selection of [0.89, 0.91] along the valence dimension in the histogram, c)vad-diagram, numerical values are displayed when hovering over that point in the graph, d)index displaying the occurence of the selected words along text. e)list of selected words}
    \label{walkthrough2}
\end{figure}

The viewer uses the vad-diagram (see Fig. \ref{walkthrough1}Ab) that displays the numerical averages of the three dimensions in three single graphs to confirm the first impression gained through the examination of the histogram and to gain additional insights about the development of emotions along the text.
The valence and the dominance graphs fully consist of averages that are larger than 0.
The arousal graph consists of averages that are all about -1.
The position of the graphs confirms the first impression that the histogram provided, namely that the text generally contains words that are generally positive, in control, and calm. In addition, the vad-diagram indicates the development of the three affective dimensions along the text. For example, valence in the beginning of the text is generally higher than towards the end. The arousal graph differs a little more along the text than the dominance graph, but not as much as the valence graph. In contrast, the dominance and arousal do not vary as much across the text.

\subsection{Comparing peculiarities in visualization to content}

\begin{figure}[th]
    \includegraphics[width=\columnwidth]{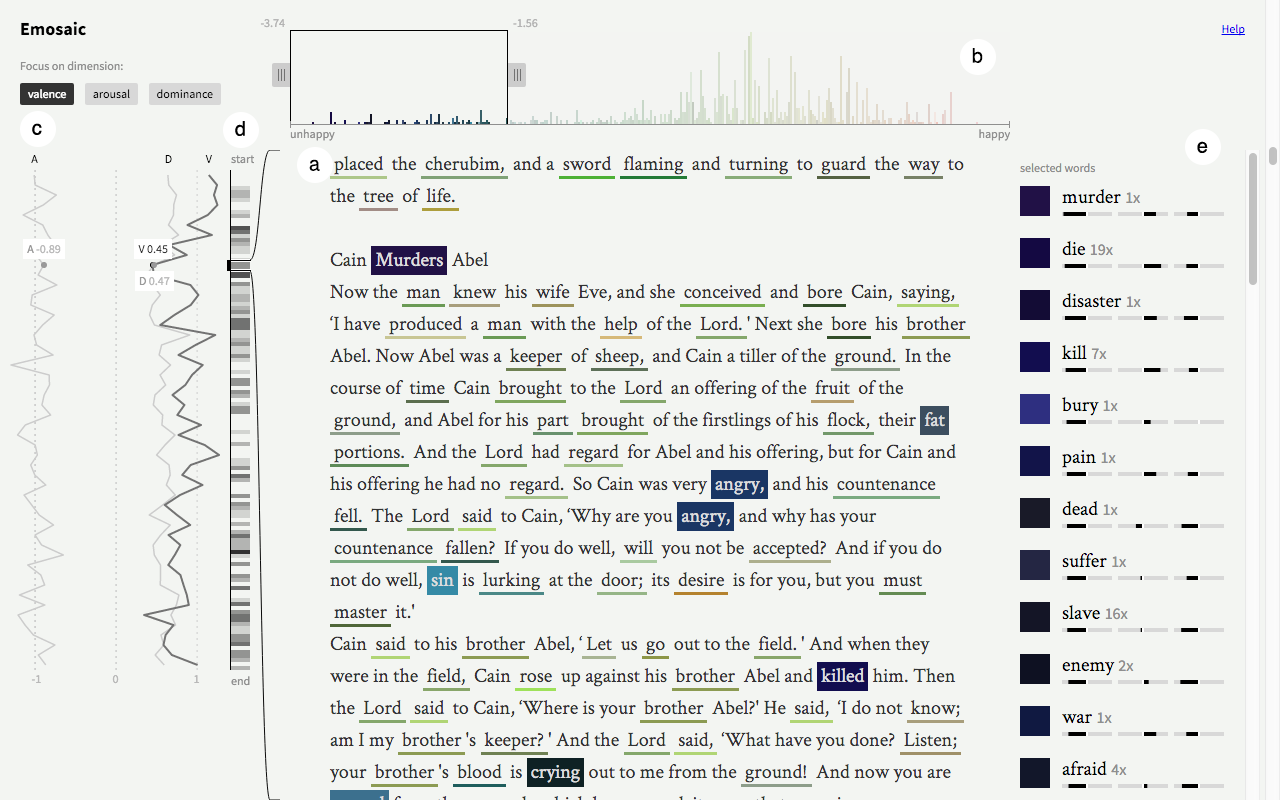}
    \caption{a) Textview, b) selection of [-3.74, -1.56] along the valence dimension in the histogram, c) vad-diagram, d) index displaying occurence of selected words, and e) micro-view with list of selected words.}
    \label{walkthrough3}
\end{figure}

Curious about the variability in the valence dimension, the scholar examines the valence more closely (see Fig.~\ref{walkthrough2}c).
Valence and dominance seem to vary only slightly in the first part of the text. By hovering over the highest peak in the valence histogram in the top part of the interface (see Fig.~\ref{walkthrough2}b), it is possible to reveal its numeric value of [0.89, 0.91]. When selecting this specific bar in the histogram, words of that valence are represented in the micro view (see Fig. 5e) and their occurrence within the text is indicated as color highlights in the text view as well as in the index (Fig.~\ref{walkthrough2}a,d). The higher the average occurrence in the text passage, the darker the grey in the index. When looking at the index the viewer sees that words within the current selection are used mostly in the first text passage. After reading the text passage, it can be stated that this is the part where God creates the earth, thus a positive story, also indicated by the high averages in the valence graph in the vad-diagram (Fig.~\ref{walkthrough2}c). Since [0.89, 0.91] is only a slightly positive rating, it does not explain the overall high rating indicated by the vad-diagram. Therefore, the reader selects a valence scope further towards the positive end [1.52, 3.53] in the histogram and the index reveals a high occurrence of positively rated words, when the same selection is moved to the negative end of the histogram to a scope of [-3.74, -1.71], the index reveals that there are few to no such negative words at the very beginning of the text (see Fig.~\ref{walkthrough3}d). This results in an insight that the positive average of the valence dimensions in this passage comes from a relatively high number of words that have been rated as happy or almost neutral tending to happy.

The scholar does not want to explore the arousal and dominance dimensions in detail. However, the slight variation of the emotional averages of dominance and valence suggests the usage of an emotionally similar vocabulary in this section. By reading the text, the scholar can state that “And God said …” are the first words of almost every sentence, indicating similar contents of the sentences of this passage. A similar vocabulary and similar semantic expressions explain the low variation of the valence graph.

Since valence seems to differ a lot more throughout the text than the other dimensions, it seems promising to examine this phenomenon in more detail. The reader hovers over a peculiar negative peak because it seems to be the longest passage throughout the text that has been rated unhappy, indicated by two following negative anchor points in the diagram (see Fig.~\ref{walkthrough3}c). The average value of all dimensions appear at the point of the diagram that is hovered over. After clicking on this particular point of the diagram, the index and the text in the text-view scroll to the same position, indicating that the position on the index always reflects the position in the text. The passage has the title “Cain Murders Abel” (see Fig.~\ref{walkthrough3}a). To further investigate this text section, the viewer selects a negative valence scope [-3.74, -1.56] in the histogram (see Fig.~\ref{walkthrough3}b). The respective words are highlighted and reveal that the passage contains “murder”, which is the word with the lowest valence score in the entire text (see Fig.~\ref{walkthrough3}e). The index view indicates through a darker grey that there is a high occurrence of negative words in this passage (see Fig.~\ref{walkthrough3}d). By examining the index and scrolling through the text and by clicking on other negative peaks in the vad-diagram, it becomes visible that there are other passages with a similar or even higher occurrence of unhappy words (see Fig.~\ref{walkthrough3}d). To examine why this passage still is rated as one of the unhappiest, the user moves the selection to a positive valence scope of [1.35, 3.52]. The index shows a very light grey, thus a low occurrence of positive words. Other passages that also have a rather negative average peak show a higher occurrence of positive words. This leads to an insight that the outstanding negative average does not only result from a high occurrence of unhappy words, but also from a relatively low occurrence of happy words.

\begin{figure}[htbp]
    \includegraphics[width=\columnwidth]{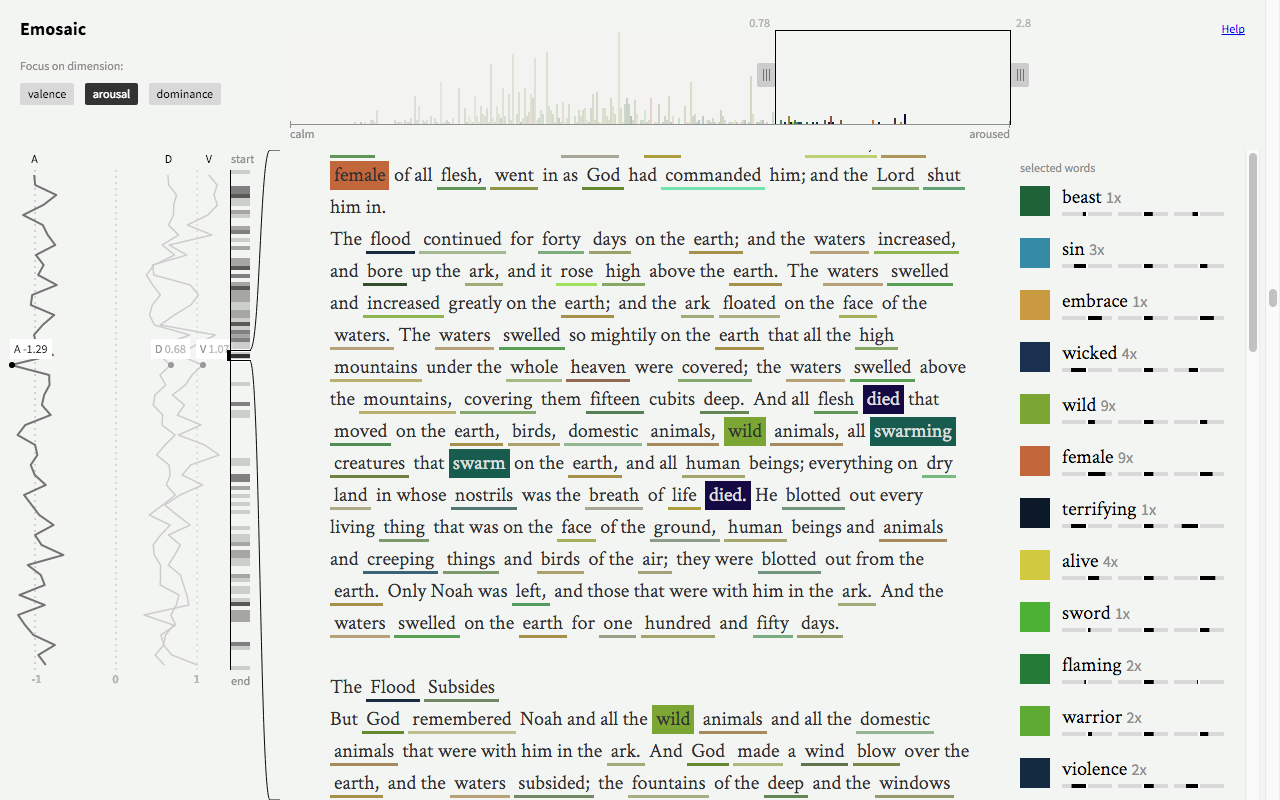}
    \caption{Visualization of words of a selected arousal scope of [0.78, 2.8] }
    \label{walkthrough4}
\end{figure}

To further examine peculiarities in the occurrence of words with high arousal scores, the reader selects a range  towards the right end of the arousal histogram [0.78, 2.8] (see Fig.~\ref{walkthrough4}). By clicking on the darkest section in the index, the text scrolls to the passage of “The Great Flood”, in which God floods the world. The content of this passage evidently corresponds with the increase of arousal. After selecting a range towards the negative end of the arousal histogram [-3.4, -1.38], the index reveals two passages where arousal seems to be particularly low (see Fig.~\ref{walkthrough4}). One is called “Judgement Pronounced on Sodom”; a close reading of this section confirms the arousal ratings to be low. The other text passage called “The Flood Subsides” however, does have a high conformity with its low ratings of arousal. It is the passage that describes the drawback of the flood.

\section{Discussion}

Our goal for the color mapping was to develop a system that can visualize three dimensions on a planar space, attempting to convey emotional content through merely one visual variable, namely color. We wanted to create an interface that provides a framework for analyzing and exploring our visualization of emotional content of text by combining distant and close reading through visualizations of different level of detail. In the following we reflect on this ambition and raise questions for future research and design of text visualizations, with particular emphasis on revealing emotion.

\subsection{Color}

After evaluating the color mapping by a walkthrough and informal discussions with digital humanities researchers, we believe that the HSV color space provides promising conditions for our purpose, but are aware of the fact that it also has its limitations. The color parameters are independent and can be relatively easily decoded by the user because of their similarity to the human perception of color. However, if brightness is very low, the other parameters do not have much impact on color perception, because the final color result would always be black, thus differentiation becomes more difficult. This could be addressed by using a non-linear mapping that takes these perceptual challenges into account, yet, this may in part counteract the purpose of a simplistic color mapping that can be easily explained and interpreted through its different color parameters. Even though we are aware of that issue, we believe that this scenario will only affect a limited number of words, namely those of the very end of the negative dominance spectrum~\cite{Warriner2013}. Since we additionally represent the numerical values for each word in the micro view, we provide a different solution for the decoding of the independent emotional parameters of these exceptions within the interface.

Another problem concerning color perception might be that equidistant colors in the HSV color space are seldom perceived as equidistant. This results out of the fact that the “value/brightness” parameter of HSV is not the measure for the actually perceived brightness, but the physical lightness. This means that the color parameters are theoretically independent, but hue and perceived brightness are interdependent. For the analysis in our tool, this is unlikely to be a problem when we try to compare tendencies of whole text passages or differences between emotionally similar words. However, if we want to compare words on the same end of the valence (hue) scale, that differ only slightly in their arousal or dominance values, equidistant perception of hues can pose a challenge. For this reason, the interface does provide the emotion parameters as numerical values.

The determination of hue in our color mapping is based on an informal hypothesis that love should be assigned to a hue of red. We do, however, know that it has been shown that other emotions are associated with red~\cite{Valdez1994}. Future versions of our prototype could integrate a custom definition of hue by the reader to enhance intuitive readability.

\subsection{Comparison}

One change in a view affects the visualizations in all other views, because all views of the interface are coordinated. This means that moving from one point of interest to another without losing the current text position is hindered, thus a comparison of different text parts to each other is difficult.
The current version of the prototype does not support the comparison of texts, since the synchronization of the views was a necessary choice to enable the display of multiple levels of detail and the combination of close and distant reading. It facilitates the exploration of the emotional content along different emotional dimensions throughout several different levels of aggregation at a glance. At the moment, comparative analysis can only be carried out by juxtaposing multiple browser windows and comparing the analytic views, especially the macro views and histograms for different texts. However, there are certainly analysis tasks that would benefit from comparative views that explicitly reveal differences and similarities of emotional content in two texts.
For the purpose of this research we focused on the maximum granularity of one article, a chapter, or a book. However, many literary and linguistic analyses focus on entire corpora such as an author’s reference library or series of books published by the same author. The challenge would be how the analysis of affective content across a collection of books can be meaningfully supported.

\subsection{Preprocessing}

The structure of sentiments and topics in written text is considerably complex. To make sense of the affective meaning of a word, aspects of the local context of a word need to be taken into account, but also negation, superlatives, comparatives, or stop-word removal. We use lemmatization in order to find entries in the database, but we know that the color mapping can be even further improved by using other natural language processing methods. We plan to include additional preprocessing methods to enhance results of the color mapping.
\section{Conclusion}

Regardless of whether it is literary work, historic source, or human communication, with the increasing amount and availability of digital text, there is an opportunity to devise new ways of making sense of written language. While distant reading visualizations enable the reader to see global patterns of a text as abstract overviews, there is a common understanding that the close-up interpretation of individual sources and sentences cannot be replaced by such high-level analysis. A particular important aspect of language is the affective component that many words are strongly associated with. This emotional layer can provide an additional analytical framework for the study of text, however, so far it has been difficult to incorporate sentiment information into the reading environment. With this research we have explored this gap and made two main contributions:

\begin{itemize}
\item A color mapping that represents positions in a three-dimensional emotional space as unique colors, and 
\item a text visualization environment revealing the emotional content of texts using a range of visual and interactive representations.
\end{itemize}

Emosaic features different visualizations along different levels of abstraction, resulting in a fully functional toolset that enables exploration of texts concerning their emotionality in combination with direct access to the underlying text. The presented visualization techniques represent text and its sentiment in a way that does not remove the original form of a text and thus enables the parallel pursuit of close and distant reading.


\bibliographystyle{abbrv}
\bibliography{emosaic}
\end{document}